# PERFORMANCE ANALYSIS OF CARRIER AGGREGATION FOR VARIOUS MOBILE NETWORK IMPLEMENTATIONS SCENARIO BASED ON SPECTRUM ALLOCATED.

Liston Kiwoli [1], Anael Sam[2] and Emmanuel Manasseh[3]

[1]The Nelson Mandela African Institute of Science and Technology,
P.O Box 447 Arusha, Tanzania

[2]The Nelson Mandela African Institute of Science and Technology,
P.O Box 447 Arusha, Tanzania

[3]Tanzania Communications Regulatory Authority

## ABSTRACT

*Carrier Aggregation (CA) is one of the Long Term Evolution Advanced (LTE-A) features that allow mobile network operators (MNO) to combine multiple component carriers (CCs) across the available spectrum to create a wider bandwidth channel for increasing the network data throughput and overall capacity. CA has a potential to enhance data rates and network performance in the downlink, uplink, or both, and it can support aggregation of frequency division duplexing (FDD) as well as time division duplexing (TDD). The technique enables the MNO to exploit fragmented spectrum allocations and can be utilized to aggregate licensed and unlicensed carrier spectrum as well.*

*This paper analyzes the performance gains and complexity level that arises from the aggregation of three inter-band component carriers (3CC) as compared to the aggregation of 2CC using a Vienna LTE System Level simulator. The results show a considerable growth in the average cell throughput when 3CC aggregations are implemented over the 2CC aggregation, at the expense of reduction in the fairness index. The reduction in the fairness index implies that, the scheduler has an increased task in resource allocations due to the added component carrier. Compensating for such decrease in the fairness index could result into scheduler design complexity. The proposed scheme can be adopted in combining various component carriers, to increase the bandwidth and hence the data rates.*

## KEYWORDS

*Carrier aggregation, LTE-Advanced, Fairness index, cell throughput, inter band Carrier Aggregation*

## 1. INTRODUCTION

Carrier Aggregation is an LTE-A feature that bonds together bands of spectrum to form wider channels which ultimately yield larger network capacity and convey faster speeds to UEs based on their categorical capabilities [1]. Basically it creates a broader lane that permits more data traffic to move at higher rates [1]. It enables MNO to provide high data rates without wide contiguous frequency band allocations, and ensures statistical multiplexing gain by distributing the traffic dynamically over multiple carriers [2]. When CA is implemented, operators can take asymmetrical bands into use with Frequency Division Duplex (FDD), as there can be uplink or downlink only frequency bands [2]. CA permits LTE to achieve the goals mandated by International Mobile Telecommunications Advanced (IMT-A) while maintaining backward compatibility with LTE release8 and 9 [3]. One of the key requirements for IMT-A as set by the International Telecommunication Union, Radio-communication Sector (ITU-R) is the support for variable bandwidths with encouragement to support up to 100 MHz of bandwidth [4]. With this capability, IMT-A supports enhanced user data service demands, with peak data rate targets of 1,000 Mbps for low mobility and 100 Mbps for high mobility [5]. With the fast growing demand for mobile data service, it becomes more and more difficult to allocate a wide and contiguous frequency bandwidth to support high speed data communication for the end-user equipment [6]. CA lets MNO support high data rates over wide bandwidths by aggregating frequency resources in the same or in different frequency bands [6]. Each aggregated carrier is known as a CC. The CC can have a bandwidth of 1.4, 3, 5, 10, 15 or 20 MHz, with a maximum of five CCs aggregation at a time, which means, the maximum aggregated bandwidth achievable at a time is 100 MHz [3].

The goal of this research is to determine the performance gains and complexity that arises from the aggregation of three inter band CC introduced in LTE release12. The parameters of interest considered in this study are the cell throughput and Fairness Index.

The rest of this paper is organized as follows: Section II provides an overview of system model while Section III covers simulation results and discussions followed by conclusion in Section VI.

## 1.1 Background

The valuable resource in wireless communication systems is spectrum which is widely focused area of research over the few decades [7]. In the 3rd Generation Partnership Project (3GPP), the radio interface specifications for the next generation mobile systems were finalized as release8 and called LTE [8]. LTE uses orthogonal frequency division multiple access (OFDMA) and single-carrier frequency division multiple access (SC-FDMA) as the multiple access scheme in the downlink and uplink, respectively [8].

OFDM is an attractive air-interface for next- generation wireless network [9]. It allows data to be distributed across a large number of carriers that are accurately spaced apart orthogonally. OFDM is well known for its excellent robustness against multipath channel and the its use of low complexity equalizers at the receiver [9]. LTE is full Internet Protocol (IP) based radio access that also incorporates the capability to support traffic with various levels of Quality of Service (QoS) [8].

In November 2007, during the World Radio-communication Conference, the radio frequency spectrum for IMT-A was decided. The IMT-A imposed new throughput requirements that were specified as LTE release10 and beyond. LTE release10 enhances the capabilities of LTE, to make the technology compliant with ITU's requirements for IMT-A, and the resulting system is known as LTE-A. LTE-A is characterized by its new features which are, CA, Enhanced Multiple-Input Multiple-Output (MIMO) and the support of Heterogeneous Networks (HetNet).

In LTE-A, it is necessary to support a wider bandwidth than that for LTE release8, (i.e. wider than the maximum of 20MHz per CC), in order to satisfy the high level requirement for the IMT-A's target peak data rate (PDR). Therefore, LTE-A supports CA of up to a maximum of five CCs, thereby attaining a maximum of 100MHz bandwidth channel [10]. The aggregated CCs are backward compatible with LTE release8, meaning that each CC appears as LTE release8 carrier toward LTE release8 User Equipment (UE) [11]. To achieve further improvements in the IMT-A's target PDR, LTE-A supports enhanced MIMO technology. Enhanced MIMO helps to improve the spectrum efficiencies to around 30b/s/Hz in the downlink and 15b/s/Hz in the uplink [12]. In LTE release8, the spectral efficiencies were 15b/s/Hz for downlink and 6.75b/s/Hz [12]. Heterogeneous network implementation aims to further improve spectral efficiency per unit area using a mixture of Macro-cell, Pico-cell, Femto-cell and relay base stations [13]. Most of these technologies are included in the LTE release10 specifications [14]. However, apart from the above measures to improve system bandwidth and spectral efficiency, spectrum sharing is envisioned as one of the viable approaches to achieve higher operational bandwidth efficiency and meet the growing mobile data traffic demand in a well-timed manner [15]. In [16] a price based spectrum sharing scheme for connection oriented traffic in wireless cellular networks was proposed. The sharing scheme resulted to both improved quality of service offered as well as the improved spectrum utilization. Also Mehdi B, in [17] proposed a hierarchical approach in spectrum sharing in which wireless competitive operators share the same spectrum band. Through simulation, it was revealed that spectrum efficiency can be improved through resource sharing between radio access networks taking interference into account. It was further shown in [17] that inter-operator spectrum sharing improves both low and high data rates.

This paper concentrates on CA, among the features and spectrum concepts introduced above. This is because of the CA potentials in creating larger 'virtual' carrier bandwidths for LTE services by combining separate spectrum allocations. Therefore this study is expected inform those MNOs who are yet to implement CA, of the potentials that CA can offer in utilizing their smaller chunks of frequency in the multiples of the standard CCs to create larger bandwidths for LTE services which will eventually result into optimal allocated spectrum utilization.

The carriers to be aggregated can be Intra‐band contiguous; which is the simplest CA deployment scenario that aggregates multiple adjacent CC in a single operating band [5]. However in most countries the spectrum allocation is more fragmented and contiguous aggregation within the same frequency band may not be possible [18]. For such scenarios, non-contiguous CA is deployed. The main advantage of LTE is that, it can be deployed in various frequency bands [19]. To serve huge customer base, MNOs have license to operate in multiple bands. To aggregate CCs from different frequency band, the MNOs deploy Inter band CA [19]. Inter‐band CA is more complex than intra‐band CA because the multi‐carrier signal cannot be treated as a single signal meaning that it requires a more advanced transceiver in the User Equipment [19]. Fig. 1 shows, the three CA configurations.

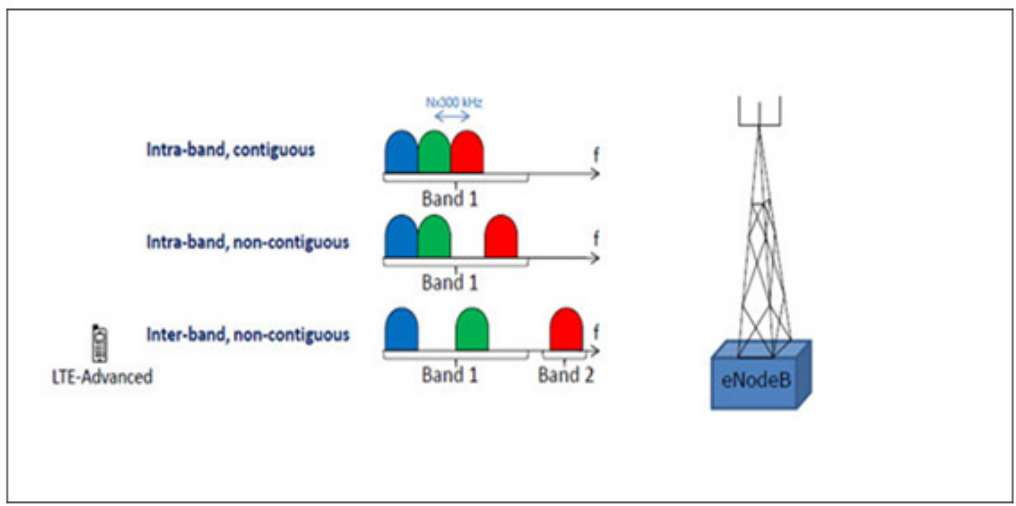

Fig 1:  CA Intra-band and Inter-band alternatives[20]

When CA is used, a number of serving cells are involved, one for each CC [20]. Primary Cell (PCell) is the cell, operating on the primary frequency, in which the UE either performs the initial radio resource control (RRC) connection establishment and re-establishment procedure or the cell indicated as the primary cell in the handover process [20]. One PCell is always active in RRC_CONNECTED mode while active Scell can be more than one [3]. The SCell is a cell operating on a secondary frequency which may be configured once an RRC connection is established and which may be used to produce additional radio resources. All PCells and SCells are known collectively as serving cells. The CCs on which the PCell and SCell are based are the Primary CC (PCC) and Secondary CC (SCC), respectively [3]. Furthermore it was revealed in [3] that a PCell is equipped with one physical downlink control channel (PDCCH) and one physical uplink control channel (PUCCH). Therefore, Physical Shared Channels (PDSCH/PUSCH) are transmitted on Pcell while only PDCCH may be transmitted on Scell [3].

In CA, a secondary cell is activated only in connected mode [21]. It was further indicated that, for Scell to be activated, a mobile device that supports Rel-10 is required to execute the generic access events that are defined for LTE Rel-8, which includes - cell search and selection, system information acquisition and initial random access. The authors in 19] claimed also, that all these procedures are carried out on the PCC for downlink and uplink. SCCs are seen as additional transmission resources which can be activated and de-activated any time depending on the capacity demands [21].

## 1.2 Related works

Since when CA was introduced in LTE release10, there have been several researches conducted to try to explore the improvements and challenges the technology brings to the MNO. In [22], performance comparison between CA and independent carrier (IC) was conducted based on LTE release10 CA where aggregation of 2CC was considered. It was reported that contiguous intra-band CA and IC systems that use the same frequency band and same total bandwidth will achieve the same performance in an LTE system as long as no additional guard band is needed to avoid interference with adjacent channels. The performance comparison between IC and CA, for other scenarios other than the contiguous intra-band CA, revealed that CA attained better performance than IC system in all networks scenarios due to the multi-user diversity and scheduling gain. Also it was established that CA with non-contiguous carriers in multiple bands could result at some inter-site distances, in a performance level which is close to the best case scenario of using contiguous spectrum allocated in relatively low frequency band. This research aims at extending the work in [22] by comparing the performance of 2CC CA and that of 3CC CA.

Three-band CA technology was enabled in LTE release12 [23]. Earlier implementations of CA enabled the use of two CCs, allowing the maximum aggregated bandwidth of 40MHz. LTE release12 enabled the use of three aggregated carriers and overall bandwidth of 60MHz [23]. Theoretically, the Evolved Node B (eNodeB) aggregates the 20MHz bandwidth carriers using CA and category 9 User Equipment (Cat9 UE) is able to achieve peak data rates of 450Mbps in the DL. In [23], the performance comparison between 2CC CA and 3CC CA was conducted. The research was based on real data collected from the live network which had already implemented both 3CC CA and 2CC CA. The tests were performed on TeliaSonera network where the performance comparison between 2CC CA and 3CC CA was conducted. Measurements for 3CC CA were done using CAT9 UE (Samsung Galaxy S6 edge plus). Galaxy S6 edge plus is capable of achieving 450Mbps in the DL using 2x2 MIMO, combining up to 60MHz of bandwidth. 2CC CA Measurements were done using CAT6 UE (Samsung Galaxy Note 4). Samsung Galaxy Note 4 is capable of achieving 300Mbps in the DL using 2x2 MIMO, combining up to 40MHz of bandwidth. All measurements were done while at stationery. In this work it was revealed that 3CC CA allows higher peak and average data rates in the DL compared to 2CC CA. This work revealed that the end user experiences a 38% growth in the average throughput when using 3CC CA as compared to 2CC CA. Also, it was revealed that 3CC CA performance was more stable compared to 2CC CA performance[23]. Although the work in[23] made comparison of the performance of 2CC CA and 3CC CA, but used proprietary data, and equipment that makes it difficult for researchers to reproduce the work for comparison and exploring more possibilities.

To address this gap, this work considers a standard LTE network simulator, the Vienna LTE system Level Simulator to quantify the performance gain attained by using 3CC CA as compared to the 2CC CA. Also, this work seeks to find out the impact 3CC CA has on the fairness index as compared to the 2CC CA which was not considered in both [22] and [23].

## 2. SYSTEM MODEL

### 2.1 Tools and Methodology

To carry out this research, the Vienna LTE System Level simulator (Vienna-LTE-A-SLS-v1-9-Q1-2016-Carrier-Aggregation) was used. The simulator is built on MATLAB. The following algorithm was used to generate data.

1. Run simulation using Vienna LTE-A System Level simulator and Matlab

   ✓ Change the values of: frequency (900MHz, 1800MHz or 2100MHz),

   ✓ The channel bandwidth (based on the standard LTE release8 CCs) and

   ✓ The CA mode (whether is 2CC or 3CC CA)

2. Record All the data from the Cell statistics Summary

3. Repeat 1 & 2 for various CC combination

4. Tabulate the recorded data

5. Visualize the collected data via MS Excel

### 2.2 System Model Parameter Descriptions

A system was set up that enabled simulation of a CA network consisting of 3CCs. In this simulation, the setup in [22] was customized where LTE CA simulation was built on top of platforms developed at the Technical University of Vienna (TU Wien). The Vienna LTE system level simulator model was implemented using MATLAB and consists of many files and functions. Some modifications were done in order to be able to run the CA Simulation. The main focus was on the average cell throughput and fairness index when comparing the CA of 2CC to the aggregation of 3CCs. These two parameters of interest will be delineated further in the following paragraphs.

### 2.3 Cell Throughput

The greatest achievement of LTE is the high data rate it can provide over a wireless channel. If we go by the Shannon-Hartley formula for channel capacity, we see that he throughput in bits per second depends on the channel bandwidth and the signal quality (SNR).

$$C = B * \log(1 + SNR) - - - - - - - - - - - (1)$$

Where,

$C$ is the channel capacity, $B$ is the channel bandwidth and **SNR** is the signal to noise ratio From equation (1), one can observe that although Bandwidth is an indispensable factor in the overall throughput, it also depends on how we use the SNR conditions for increasing throughput. For example, higher SNR (good channel conditions) means, a higher order modulation scheme such as 64QAM can be used. As the channel conditions deteriorates (for instance fading, due to increased distance away from the serving eNodeB), lower order modulation schemes are preferred as shown below.

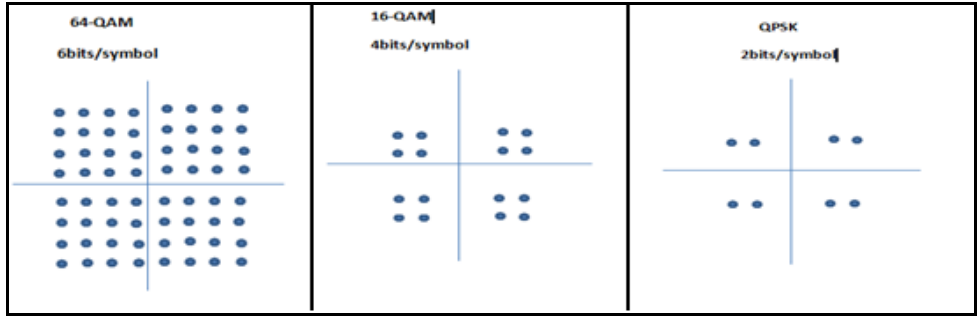

Fig 3. Modulation schemes used with LTE and the number of bits mapped to each scheme [24]

Based on the modulation scheme, the bits are mapped on the resource element. These Resource elements in turn aggregate into the Resource Block (RB). We know that in LTE we can have variable bandwidth and the higher the bandwidth the more the number of Resource Blocks. Table 1, below shows the maximum number of RBs per channel bandwidth.

Table 1.  RBs per Channel Bandwidth

| LTE Bandwidth [MHz] | Resource Block [RB] |
|---|---|
| 1.4 | 6 |
| 3 | 15 |
| 5 | 25 |
| 10 | 50 |
| 15 | 75 |
| 20 | 100 |

LTE also takes advantage of the MIMO to improve the user throughput. For instance, if all other conditions are kept the same, the channel capacity can double with a 2x2 MIMO. In our simulation the MIMO settings were 2x2, meaning two transmit antenna at the BS and 2 receiver antenna at the UE. We decided to use 2x2 MIMO in order to have fair comparison with Tanner's work in [23]. Therefore, we assumed that, with MIMO setting and other parameters kept constant (as will be explained below), cell throughput is expected to increase with the increase in channel bandwidth through CA. And 3CC CA is expected to bring better results than 2CC CA.

## 2.4 Fairness

In [25], it was reported that fairness is an important performance criterion in all resource allocation schemes. Quantitatively, fairness in resource allocation is measured by the index of fairness. This index is bound between 0 and 1 and is applicable to any resource sharing or allocation problem. This index measures the 'equality' of user's allocation in multi-user resource allocation. In this research, Proportional Fair scheduling was employed where the index is bound between 0 and 1, with 1 which is equivalent to 100% indicating that the scheduler allocated equal share to all users regardless of the measured CQI and 0 which is equivalent to 0% indicating unfair allocation, where all the available resources are allocated to few users while others get nothing [26].

In general, there is a trade-off between performance and fairness when designing a scheduling algorithm. For example, an algorithm that simply maximizes throughput would assign most of the resource blocks to devices that are closest to a cell tower, and thus can get a higher data rate with any given resource block, while devices at the edge of the cell may starve. This disparity could be even greater in an inter-band scenario, where devices close to the cell tower dominate use of the low-frequency bands [22]. For this simulation, we choose to apply the well-studied Proportional Fair (PF) scheduling algorithm [26]. With this scheduling algorithm users with better channel quality will have a higher average throughput [22]. The fairness index will be examined in this research, when a 2CC CA will be compared to 3CC CA.

## 2.5 Simulation model assumptions and inputs

Table 2 shows, the main input parameters related to frequency. All are constant inputs except the number of fragments, where in this research we expect to vary the number of fragments and examine the improvement in terms of user throughput and fairness, while also examining the increase in complexity level.

Table 2. Main input parameters

| MAIN INPUTS RELATED TO FREQUENCY BANDS | |
|---|---|
| **Input** | **Value** |
| Total bandwidth | 5 - 80 MHz |
| Block bandwidth | 1.4, 3 and 5 MHZ |
| Number of fragments | From 1 to 5 blocks |
| Transmit Power | 43 dBm for 1.4, 3 & 5MHz bandwidth carriers |
| | 46 dBm for 10 and 20 MHz bandwidth carriers |
| Frequency bands | 900MHZ, 1800MHZ and 2100MHZ |
| Antenna Gain | 16dBi (900MHZ), 18dBi (1800&2100MHZ) |
| Path Loss Model | Okumura-Hata (for 900MHZ) and COST 231 -Hata (for 1800&2100MHZ) |

Table 3 shows, the miscellaneous input parameters and settings required for simulating CA using Vienna LTE System Level Simulator. In this simulation we considered CC of different bandwidth in both cases. Case I: carrier aggregation of 2CC and Case II: carrier aggregation of 3CC.

Table 3.  Simulation Parameters

| SIMULATION PARAMETERS | |
|---|---|
| **nput** | **Value** |
| Antenna Configuration | 2x2 MIMO |
| Transmission  Mode | CLSM (Closed Loop Spatial Multiplexing) |
| TTI Length | 1ms |
| Simulation time | 20 TTI |
| RB Bandwidth | 180kHz |
| UE Distribution | Uniform |
| Antenna Azimuth Offset | 30 |
| Antenna Down tilt | 8 |
| Feedback Channel delay | 3 TTI |
| SINR averaging algorithm | MIESM |
| Sectors per cell | 3 |
| UE antenna gain | 0dB |
| Channel Trace length | 5s |
| Coupling loss | 70dB |
| UE speed | 5/3.6km/h |
| Site height | 20m |
| Receiver height Site height | 1.5m |

## 2.6 SIMULATION RESULTS AND DISCUSSIONS

This work tries to find out which CC combination can result into better average cell throughput. Fig. 4 shows, the comparison between various 2CC CA combinations. It was found that a combination of CCs from 1800MHz and 2100MHz yield better average cell throughput than the other two component carrier aggregations. Among other reasons, this is because with higher frequencies, we have a possibility of higher bandwidths. This is well indicated in [27], when you look at band 1, 3 and 8 which were considered in this research work.

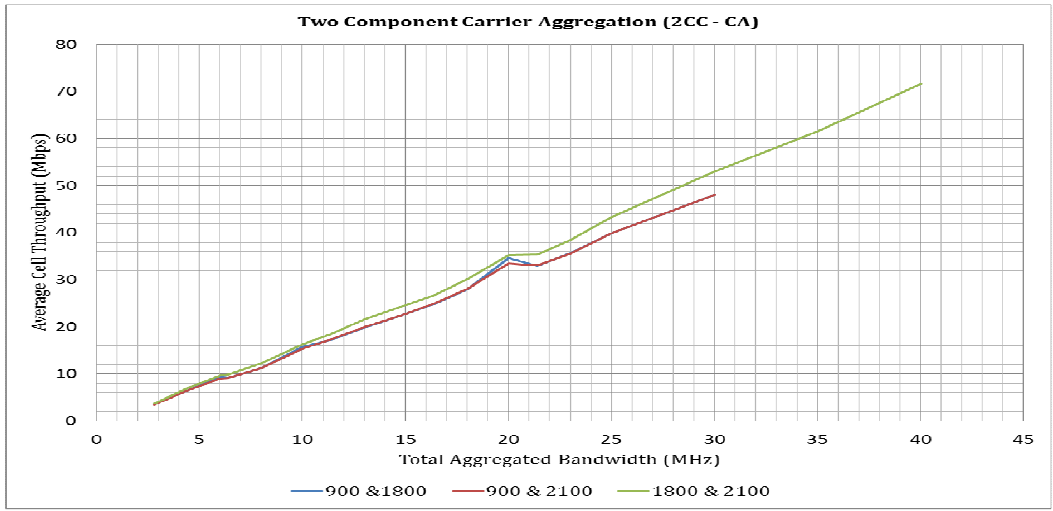

Fig 4: 2CC Carrier aggregation Throughput Comparison

By using the Vienna LTE system level simulator (v1-9-Q1-2016), a maximum of 71.6Mbps cell average throughput was obtained. This was obtained when 20MHz bandwidth of 1800MHz frequency was aggregated with 20MHz bandwidth of 2100MHz frequency. Other than this combination, all 2CC CA yielded less than 71.6Mbps cell average. The maximum aggregated bandwidth that was logically possible for 2CC CA was 40MHz.

The 3CC CA enabled a maximum of 50MHz aggregated channel bandwidth with a maximum of 10MHz channel bandwidth for 900MHz frequency, 20MHz channel bandwidth for 1800MHz and 20MHz channel bandwidth for 2100MHz. Using the Vienna LTE system Level Simulator stated above, a maximum of 79.93Mbps average throughput was obtained. Comparing maximum average throughput obtained from the 2CC CA, and that obtained in 3CC CA, data rate increment of 11.6% was realized.

In Fig. 5 shows, the fairness index variations for the 2CC inter band CAs. A quick observation shows that, the highest fairness index values were obtained when the two CCs are equal in bandwidth. This implies that the scheduler allocates nearly equal resources to the 2CC. Another observation is that, when the CCs in the inter-band CA are equal, the CA between 900MHz and 1800MHz gives the best fairness index, otherwise the inter band CA between 1800MHz and 2100MHz had the best fairness index values.

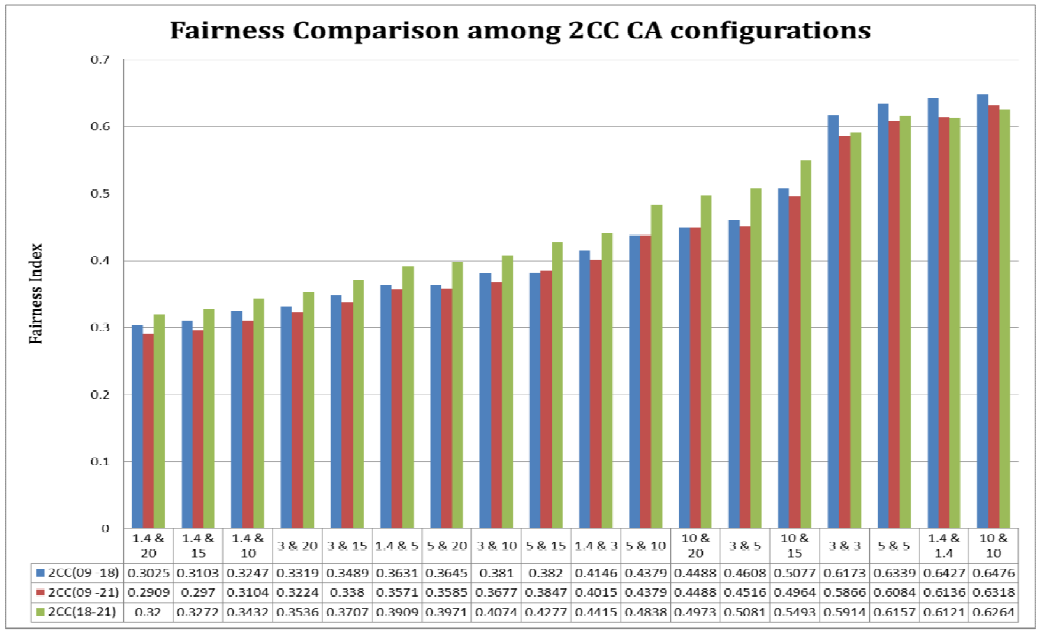

| | 1.4 & 20 | 1.4 & 15 | 1.4 & 10 | 3 & 20 | 3 & 15 | 1.4 & 5 | 5 & 20 | 5 & 10 | 3 & 10 | 5 & 15 | 1.4 & 3 | 5 & 5 | 10 & 20 | 3 & 5 | 5 & 10 | 10 & 15 | 3 & 3 | 5 & 5 | 1.4 & 1.4 | 10 & 10 |
|---|---|---|---|---|---|---|---|---|---|---|---|---|---|---|---|---|---|---|---|---|
| 2CC(09-18) | 0.3025 | 0.3103 | 0.3247 | 0.3319 | 0.3489 | 0.3631 | 0.3645 | 0.3645 | 0.381 | 0.382 | 0.4146 | 0.4379 | 0.4488 | 0.4608 | 0.5077 | 0.5173 | 0.6173 | 0.6339 | 0.6427 | 0.6476 |
| 2CC(09-21) | 0.2909 | 0.297 | 0.3104 | 0.3224 | 0.338 | 0.3571 | 0.3585 | 0.3677 | 0.3847 | 0.4015 | 0.4379 | 0.4488 | 0.4516 | 0.4964 | 0.4973 | 0.5081 | 0.5866 | 0.5914 | 0.6084 | 0.6136 |
| 2CC(18-21) | 0.32 | 0.3432 | 0.3536 | 0.3707 | 0.3909 | 0.3971 | 0.4074 | 0.4277 | 0.4415 | 0.4838 | 0.4379 | 0.5081 | 0.5493 | 0.5493 | 0.5914 | 0.6157 | 0.6121 | 0.6264 | | |

Fig 5: 2CC Carrier aggregation Fairness index Comparison

From the above observation, the value of fairness index largely depends on how much the CCs are close to each other in values; it was anticipated that in the 3CC CA, the fairness index will be lower than that of 2CC CA. Fig 6 shows, the same expected results.

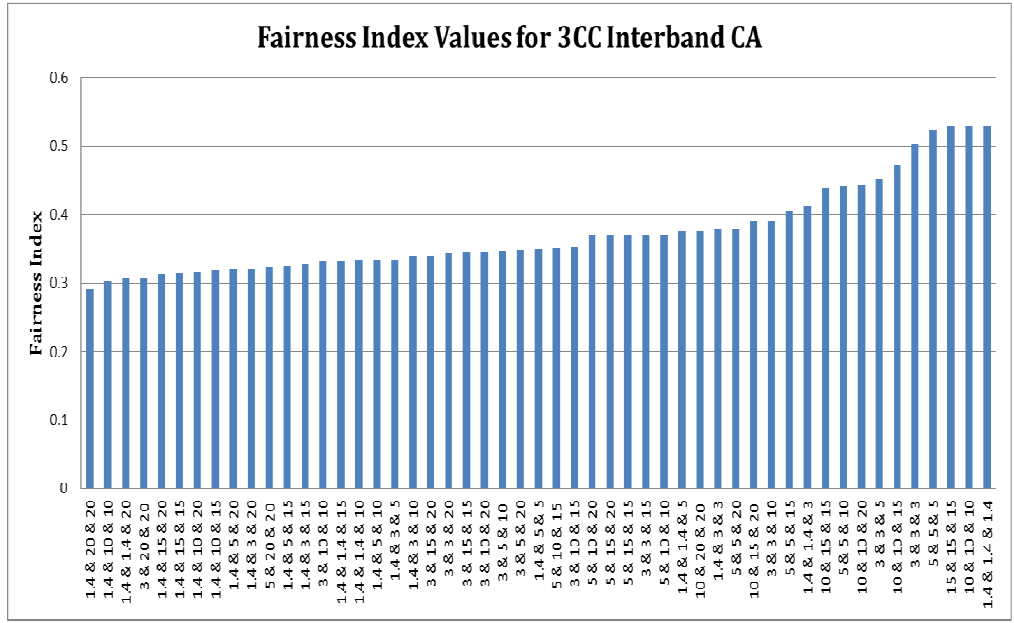

Fig 6: 3CC Carrier aggregation Fairness index values

Finding from the study shows that there is 11.6% growth in average cell throughput when using 3CC CA as compared to the 2CC CA. Although the gain is much smaller compared to 38% growth reported in [23], the reason for this could be the vendor related parameter tuning done by TeliaSonera during 3CC CA implementation and the differences in the methodology used. The proposed approach use a standard tool to conduct the study so as to enable other researchers to reproduce easily the study and dig deeper for validation of the study or for other break through, something that seemed difficult by using the methodology used in [23].

Further, it was found that, there was 18.3% decrease in the fairness index which suggests that, the fairness in scheduling resources among users based on PF scheduling would be impacted by the increase of a CC. In both cases of 2CC CA and 3CC CA, the best values for Fairness Index were obtained when the CC bandwidth were equal. And the minimum was obtained when the difference in bandwidth between the CCs was the largest. To compensate this decrease in Fairness index (if possible) could therefore result to scheduling design complexity.

## 3. CONCLUSION

In this paper, a simulation of 2CC and 3CC CA was done using the Vienna LTE System Level simulator, and the simulation results revealed that there is 11.6% increase in the average cell throughput using 3CC CA as compared to 2CC CA. However, this gain in data throughput came at the expense of 18.3% reduction in fairness index whose compensation could result into scheduler design complexity. The results also shows that the best CC combination in an inter-band 2CC CA was when CCs from 1800MHz frequency and 2100MHz were aggregated. The results are based on data analyzed when the simulations were done for radio frequencies in the following bands, 900MHz, 1800MHz and 2100MHz. In the future works, simulation can be done for more frequency bands to validate the results as well as examining the performance of other parameters which were not considered in this study such as cell edge UE throughput performance.

## REFERENCES

[1]     A. NORTON, "SPRINT AND SAMSUNG TAKE THREE-CHANNEL CARRIER AGGREGATION INTO THE FIELD WITH LIVE DEMONSTRATION IN CHICAGO | SPRINT NEWSROOM," HTTP://NEWSROOM.SPRINT.COM/,2016.[ONLINE].AVAILABLE: HTTP://NEWSROOM.SPRINT.COM/SPRINT-AND-SAMSUNG-TAKE-THREE-CHANNEL-CARRIER-AGGREGATION-INTO-THE-FIELD-WITH-LIVE-DEMONSTRATION-IN-CHICAGO.HTM. [ACCESSED: 16-AUG-2017].

[2]     NOKIA, "WHITE PAPER -NOKIA SOLUTIONS AND NETWORKS 2014," PP. 1–27, 2014.

[3]     ANRITSU, "UNDERSTANDING LTE-ADVANCED."

[4]     K. ETEMAD, I. CORPORATION, M. FONG, R. NORY, AND R. LOVE, "CARRIER AGGREGATION FRAMEWORK IN 3GPP LTE-ADVANCED," IEEE COMMUN. MAG., VOL. 48, NO. AUGUST, PP. 60–67, 2010.

[5]     M. A. M. AL-SHIBLY, M. H. HABAEBI, AND J. CHEBIL, "CARRIER AGGREGATION IN LONG TERM EVOLUTION-ADVANCED," PROC. - 2012 IEEE CONTROL SYST. GRAD. RES. COLLOQUIUM, ICSGRC 2012, NO. SEPTEMBER 2014, PP. 154–159, 2012.

[6]     Y. XIAO, Z. HAN, C. YUEN, AND L. A. DA SILVA, "CARRIER AGGREGATION BETWEEN OPERATORS IN NEXT GENERATION CELLULAR NETWORKS: A STABLE ROOMMATE MARKET," IEEE TRANS. WIREL. COMMUN., VOL. 15, NO. 1, PP. 633–650, 2016.

[7]     J. D. J AND R. MURUGESH, "A STUDY ON QUANTITATIVE PARAMETERS OF SPECTRUM HANDOFF IN COGNITIVE RADIO NETWORKS," INT. J. WIREL. MOB. NETWORKS, VOL. 9, NO. 1, PP. 31–38, 2017.

[8]     Y. Kakishima, T. Kawamura, Y. Kishiyama, H. Taoka, and T. Nakamura, "Experimental evaluation on throughput performance of asymmetric carrier aggregation in LTE-advanced," IEEE Veh. Technol. Conf., 2011.

[9]     N. N. and B. Ridha and Innov'Com, "Dynamic optimization of overlap-and-add length over MIMO MBOFDM system based on SNR and CIR estimate.," IJWMN, vol. 7, no. 3, pp. 1–87, 2015.

[10]    I. Shayea, M. Ismail, R. Nordin,  a Concept, and C. Aggregation, "Capacity Evaluation of Carrier Agreggation Techniques in LTE-Advanced System," no. July, pp. 3–5, 2012.

[11]    L. Chen, W. Chen, X. Zhang, and D. Yang, "Analysis and simulation for spectrum aggregation in LTE-advanced system," IEEE Veh. Technol. Conf., 2009.

[12]    S. Kasapovic, S. Mujkic, and S. Mujacic, "Enhanced MIMO influence on LTE-Advanced network performances," Elektron. ir Elektrotechnika, vol. 22, no. 1, pp. 81–86, 2016.

[13]    H. Q. Ngo, H. A. Suraweera, M. Matthaiou, and E. G. Larsson, "LTE Advanced : Heterogeneous Networks," IEEE J. Sel. Areas Commun., vol. 32, no. 9, pp. 1721–1737, 2014.

[14]    T. Specification, "Etsi ts 136 201," vol. 0, pp. 0–14, 2011.

[15]    B. Singh, K. Koufos, O. Tirkkonen, and R. Berry, "Co-primary inter-operator spectrum sharing over a limited spectrum pool using repeated games," IEEE Int. Conf. Commun., vol. 2015–Septe, pp. 1494–1499, 2015.

[16]    A. Sam, D. Machuve, and J. Anatory, "A Price Based Spectrum Sharing Scheme in Wireless Cellular Networks," vol. 3, no. 8, pp. 1–8, 2013.

[17]    M. Bennis, Spectrum sharing for future mobile cellular systems. 2009.

[18]    A. Z. Yonis, M. F. L. Abdullah, and M. F. Ghanim, "Effective Carrier Aggregation on the LTE-Advanced Systems," Int. J. Adv. Sci. Technol., vol. 41, no. April 2012, pp. 15–26, 2012.

[19]    Youtube, "9 - Carrier Aggregation Technique (CA) - Capacity & Coverage Enhancement - Fundamentals of 4G (LTE) - YouTube," https://www.youtube.com, 2017. [Online]. Available: https://www.youtube.com/watch?v=tZ50CSADKIY&t=61s. [Accessed: 31-Aug-2017].

[20]    J. Wannstrom, "Carrier Aggregation explained," 3GPP, 2013.

[21]    Rohde&Schwarz, "Carrier aggregation – (one) key enabler for LTE-Advanced."

[22]    M. Alotaibi, J. M. Peha, and M. A. Sirbu, "Impact of Spectrum Aggregation Technology and Spectrum Allocation on Cellular Network Performance," IEEE Conf. Dyn. Spectr. Access Networks, pp. 326–335, 2015.

[23]    A. Tanner, "LTE-A 3CC Carrier Aggregation," no. May, 2016.

[24]    Feedburner, "LTE Signaling: Troubleshooting and Optimization: Link Adaptation in LTE,"http://ltesignaling.blogspot.com,2012.[Online].Available: http://ltesignaling.blogspot.com/2012/02/link-adaptation-in-lte.html. [Accessed: 21-Aug-2017].

[25]   R. Jain, D.-M. Chiu, and W. R. Hawe, "A quantitative measure of fairness and discrimination for resource allocation in shared computer system," DEC technical report TR301, vol. cs.NI/9809, no. DEC-TR-301. pp. 1–38, 1984.

[26]   R. Proportional, "Reduced-Complexity Proportional OFDMA," vol. 0, no. 60472070, pp. 1221–1225, 2006.

[27]   Niviuk, "LTE Carrier Aggregation," http://niviuk.free.fr, 2017. [Online]. Available: http://niviuk.free.fr/lte_ca_band.php. [Accessed: 07-Oct-2017].

## Authors

Liston Kiwoli, Master's student at the Nelson Mandela African Institute of Science and Technology. He received his Bachelor's degree in Telecommunications Engineering at the University of Dar es salaam in 2008. Since 2008, he has worked as a Telecommunications Engineer at various mobile operators in Tanzania.

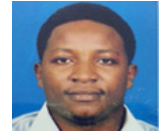

Dr. Anael Sam, Seniour Lecturer at the School of Computational and Communications Science and Engineering at the Nelson Mandela African Institute of Science and Technology from 2012 to date. From 2010 to 2012 he worked as Seniour QA (system test) Engineer at the Siemens PSE.

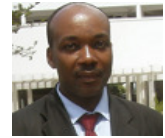

Dr. Emmanuel Manasseh, Principle Research officer at Tanzania Communications Regulatory Authority from November 2015 to date. From 2013 to March 2014, he was an assistant Professor at Hiroshima University in the Department of System Cybernetics.

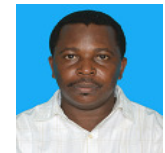